\documentclass[sigconf]{acmart}

\usepackage{latexsym}
\usepackage{graphicx}
\usepackage{amsfonts,amsmath}
\usepackage{nccmath}
\usepackage{optidef}
\usepackage{hyperref}
\usepackage{color,soul}
\usepackage{array}
\usepackage{algorithm}
\usepackage{algorithmic}
\usepackage[T1]{fontenc}
\usepackage[utf8]{inputenc}
\usepackage{subcaption}
\usepackage{multicol}
\usepackage{multirow}
\newcolumntype{L}{>{\centering\arraybackslash}m{2.2cm}}
\newcolumntype{M}{>{\centering\arraybackslash}m{3cm}}
\newcolumntype{N}{>{\centering\arraybackslash}m{1cm}}
\newcolumntype{O}{>{\centering\arraybackslash}m{2cm}}
\newcolumntype{P}{>{\centering\arraybackslash}m{1.7cm}}
\newcolumntype{Q}{>{\centering\arraybackslash}m{1.8cm}}
\usepackage{balance}
\hypersetup{draft}
\usepackage{fancyhdr}

\fancypagestyle{firststyle}{
	\fancyhf{}
	\fancyhead[L]{H.Poddar, T. Yoshimura, M. Pagin, T. S. Rappaport, A. Ishii and M. Zorzi, ``ns-3 Implementation of Sub-Terahertz and Millimeter Wave Drop-based NYU Channel Model (NYUSIM)'' \textit{in 2023 Workshop on ns-3 (WNS3 2023),} Arlington, VA, USA, June 28–29, 2023.}

}

\AtBeginDocument{%
  \providecommand\BibTeX{{%
    \normalfont B\kern-0.5em{\scshape i\kern-0.25em b}\kern-0.8em\TeX}}}

\copyrightyear{2023}
\acmYear{2023}
\setcopyright{acmlicensed}\acmConference[WNS3 2023]{2023 Workshop on ns-3}{June 28--29, 2023}{Arlington, VA, USA}
\acmBooktitle{2023 Workshop on ns-3 (WNS3 2023), June 28--29, 2023, Arlington, VA, USA}
\acmPrice{15.00}
\acmDOI{10.1145/3592149.3592155}
\acmISBN{979-8-4007-0747-6/23/06}

\begin{document}

\title{ns-3 Implementation of Sub-Terahertz and Millimeter Wave Drop-based NYU Channel Model (NYUSIM)}

\flushbottom
\setlength{\parskip}{0ex plus0.1ex}
\addtolength{\skip\footins}{-0.2pc plus 40pt}

\author{Hitesh Poddar}
\affiliation{%
  \institution{NYU WIRELESS, NYU}
  \city{New York}
  \country{USA}
}
\email{hiteshp@nyu.edu}

\author{Tomoki Yoshimura}
\affiliation{%
  \institution{SHARP Corporation}
  \city{Washington}
  \country{USA}
}
\email{yoshimurat@sharplabs.com}

\author{Matteo Pagin}
\affiliation{%
  \institution{University of Padova}
  \city{Padova}
  \country{Italy}
}
\email{paginmatte@dei.unipd.it}

\author{Theodore S. Rappaport}
\affiliation{%
  \institution{NYU WIRELESS, NYU}
  \city{New York}
  \country{USA}
}

\email{tsr@nyu.edu}
\author{Art Ishii}
\affiliation{%
  \institution{SHARP Corporation}
  \city{Washington}
  \country{USA}
}
\email{ishiia@sharplabs.com}

\author{Michele Zorzi}
\affiliation{%
  \institution{University of Padova}
  \city{Padova}
  \country{Italy}
}
\email{zorzi@dei.unipd.it}

\renewcommand{\shortauthors}{H. Poddar et al.}

\begin{abstract}
The next generation of wireless networks will use sub-THz frequencies alongside mmWave frequencies to enable multi-Gbps and low latency applications. To enable different verticals and use cases, engineers must take a holistic approach to build, analyze, and study different parts of the network and the interplay among the lower and higher layers of the protocol stack. It is of paramount importance to accurately characterize the radio propagation in diverse scenarios such as urban microcell (UMi), urban macrocell (UMa), rural macrocell (RMa), indoor hotspot (InH), and indoor factory (InF) for a wide range of frequencies. The 3GPP statistical channel model (SCM) is oversimplified and restricted to the frequency range of 0.5-100 GHz. Thus, to overcome these limitations, this paper presents a detailed implementation of the drop-based NYU channel model (NYUSIM) for the frequency range of 0.5–150 GHz for the UMi, UMa, RMa, InH, and InF scenarios. 
NYUSIM allows researchers to design and evaluate new algorithms and protocols for future sub-THz wireless networks in ns-3.
\end{abstract}

\ccsdesc[100]{Networks~Network simulation; Mobile networks;}

\keywords{3GPP, 5G, 6G, B5G, channel model, system simulation, sub-THz}

\maketitle
\thispagestyle{firststyle}

\section{Introduction}
The increasing demand for higher capacity, multi-Gbps data rates, and latencies of less than a millisecond is pushing the wireless industry to explore frequencies higher than mmWave \cite{rappaport2019wireless,jiang2021road}. The Federal Communications Commission (FCC) Office of Engineering and Technology in the ET-Docket 18-21, released in 2019, offered experimental licenses from 95 GHz to 3 THz to encourage the development of new wireless technologies for data-intensive and high-bandwidth applications \cite{FCC}. Standardization bodies and researchers worldwide have started exploring the D-band (130–175 GHz) centered around 140 GHz as a promising band for future wireless networks \cite{singh2019beyond,NextGAlliance}. With new technologies such as software-defined networking, multi-user communication networks have become too complex for traditional analytical methods to provide valuable insights and understanding of network performance. Hence, researchers rely on detailed network simulators, such as ns-3 and Monte Carlo simulations \cite{tranter2004principles,shakkottai2003cross}, to understand the overall end-to-end cross-layer network behavior. Simulations provide a quick, accessible, efficient, and accurate approach to designing and evaluating the performance of a complex wireless network. The mmWave \cite{mezzavilla2018end} and NR \cite{patriciello2019e2e} modules in ns-3 are used for full-stack end-to-end simulations of 5G wireless network. 
A network's performance depends in part on how accurately the wireless channel between the multiple next-generation Node B (gNBs) and user equipment (UE) is modeled. Early work showed that a measurement-based statistical channel impulse response (CIR) model (SIRCIM) \cite{rappaport1991statistical} could be used for more accurate cross-layer end-to-end bit error rate (BER) simulation (BERSIM) \cite{fung1993bit,rappaport1993computer} as it was found that BER was not only a function of root mean square (RMS) delay spread of the propagation channel but also of the temporal and spatial distribution of MP components. Thus, it is vital to use channel models that faithfully reproduce the real-world wireless channel across a wide range of frequencies and propagation scenarios \cite{Rap2017itap,rappaport2017investigation} to evaluate and analyze the performance of a wireless network.
\par Currently, ns-3 uses the 3GPP TR 38.901 SCM \cite{zugno2020implementation,3GPPTR} to model the wireless channel for the frequency range of 0.5-100 GHz for all 3GPP-listed scenarios (UMi, UMa, RMa, InH, and InF). Work in \cite{rappaport2017investigation} shows that 3GPP SCM provides an oversimplification of the actual wireless channel in outdoor scenarios. In addition, limited frequency range of the 3GPP SCM does not allow researches to study future networks above 100 GHz \cite{rappaport2019wireless}. Furthermore, there needs to be more understanding of how the wireless channel model impacts overall network performance 
\begin{figure*}[h!]
\centerline{\includegraphics[scale=0.48]{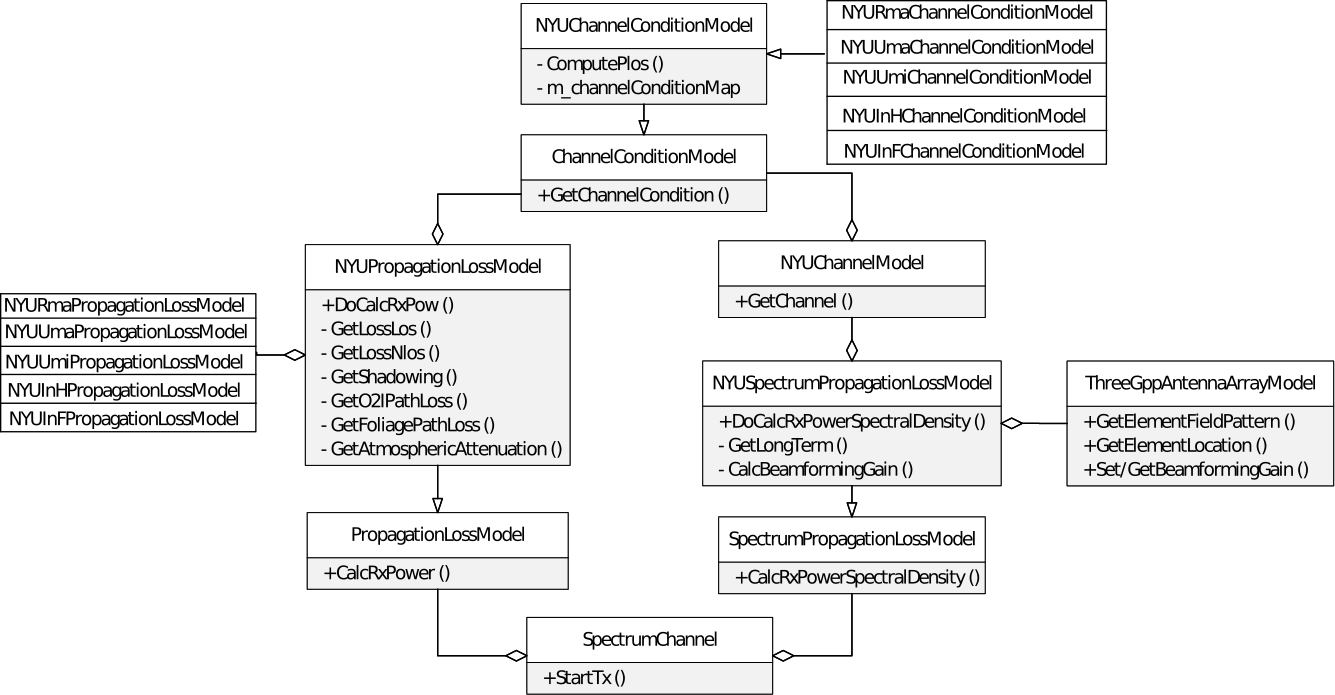}}
\caption{Simplified Unified Modeling Language (UML) Diagram for NYUSIM Implementation in ns-3}
\label{fig:UML}
\end{figure*}
\cite{hp2023icc,rappaport2017investigation,shakkottai2003cross} as ns-3 users are restricted to using only the 3GPP SCM. To enable the research community to explore and analyze networks of the future based on real-world channel measurement-based models and cover the frequency range of 0.5-150 GHz in 3GPP-listed scenarios, we present the implementation of drop-based NYUSIM in ns-3. The implementation of NYUSIM in ns-3 relies on a MATLAB-based open-source mmWave and sub-THz channel simulator, \textit{NYUSIM}, that was developed in 2016 by NYU WIRELESS. Numerous cohorts of graduate students have provided support for \textit{NYUSIM}, and the software has garnered over 100,000 downloads from wireless industry and academic communities worldwide \cite{nyusim,sun2017novel,ju2019millimeter} and is popularly used as an alternative to 3GPP SCM. \textit{NYUSIM} was created using extensive field measurements in New York City and rural Virginia and over 2 Terabytes of
measurement data was collected from 28 to 140\footnote{The radio channel propagation measurements were conducted at a frequency of 142 GHz. However, for simplicity, we interchangeably refer to this frequency as 140 GHz.} GHz during 2011-2022 for all the 3GPP-listed scenarios~\cite{rappaport2013millimeter,rappaport2015wideband,samimi20163,samimi2016local,sun2016investigation,sun2015synthesizing,maccartney2015indoor,maccartney2016millimeter,maccartney2017rural,ju20203,ju2021millimeter,ju2022sub,ju:2023:icc}. 
\par The rest of the article is organized as follows. Section 2 presents the implementation details of drop-based NYUSIM in ns-3 and provides a 3GPP-like channel generation procedure for NYUSIM. Section 3 lists examples of using NYUSIM in ns-3, and finally, in Section 4, we conclude the paper and present future research possibilities.
\section{NYUSIM implementation in ns-3}
There are multiple ways to model multi-user UE performance as UEs move in an environment. Some involve motion over a local area (small-scale channel variations), others look at point locations and consider large-scale channel effects over distances where users are dropped (drop-based), and yet other models consider the change of the channel over tracks or very small regions (spatial consistency). \textit{NYUSIM} based on MATLAB can simulate drop-based and spatial consistency-based channel simulations  with human blockage \cite{nyusim,sun2017novel,ju2019millimeter}. However, system-level simulations in ns-3 consisting of multiple UEs and gNBs, each with detailed protocol stacks, are computationally intensive. Thus, for simplicity and computational efficiency for simulations performed in ns-3, we have implemented drop-based \textit{NYUSIM} in ns-3 without human blockage \cite{NYUSIM_in_ns3}. \par In ns-3, the wireless channel is modeled using the \texttt{Prop\-agation} and \texttt{Spectrum} modules \cite{zugno2020implementation}. The \texttt{Propag\-ation} module defines the \texttt{Pro\-paga\-tionLossModel} interface and can be extended further to include different propagation models to model the large-scale fading. On the other hand, the \texttt{Spectrum} module defines the \texttt{Spectrum\-Prop\-agation\-LossModel} interface. The \texttt{Spe\-ctr\-um} module uses Power Spectral Density (PSD) to model the wireless channel's small-scale fading, i.e., fast fading and frequency selective fading. 
The implementation of NYUSIM in ns-3 conveniently follows the implementation of 3GPP SCM \cite{zugno2020implementation} to minimize the code changes required by reusing the modules of 3GPP SCM in ns-3. Figure \ref{fig:UML} shows the simplified Unified Modeling Language (UML) diagram for the classes created for implementing NYUSIM in ns-3 and the relationship among these classes. From Figure \ref{fig:UML}, we can observe that NYUSIM implementation is comprised of three main components: (i) \texttt{NYU\-Channel\-Condition\-Model}, which is used to determine the LOS/NLOS channel condition; (ii) \texttt{NYU\-Propagation\-Loss\-Model} models the large-scale fading, i.e., path loss and shadowing; and (iii) \texttt{NYU\-Spectrum\-Propagation\-Loss\-Model}, which computes the small-scale fading, i.e., fast fading and frequency selective fading. Also, note that we reuse the \texttt{Three\-GppAntenna\-ArrayModel} \cite{rebato2018multi} in ns-3 for NYUSIM. 

\subsection{LOS Probability Models}
\label{losModels}
The LOS probability models are used to statistically determine whether a device is in LOS or NLOS from the gNB. The NYUSIM LOS probability models depend on the 2D Tx–Rx separation distance and is frequency-independent, as it is solely based on the geometry and layout of an environment or scenario since diffraction may be largely ignored at mmWave and sub-THz frequencies \cite{Rap2017itap,7070688,rappaport2019wireless}. 
\begin{figure*}[h!]
\centerline{\includegraphics[scale=0.5]{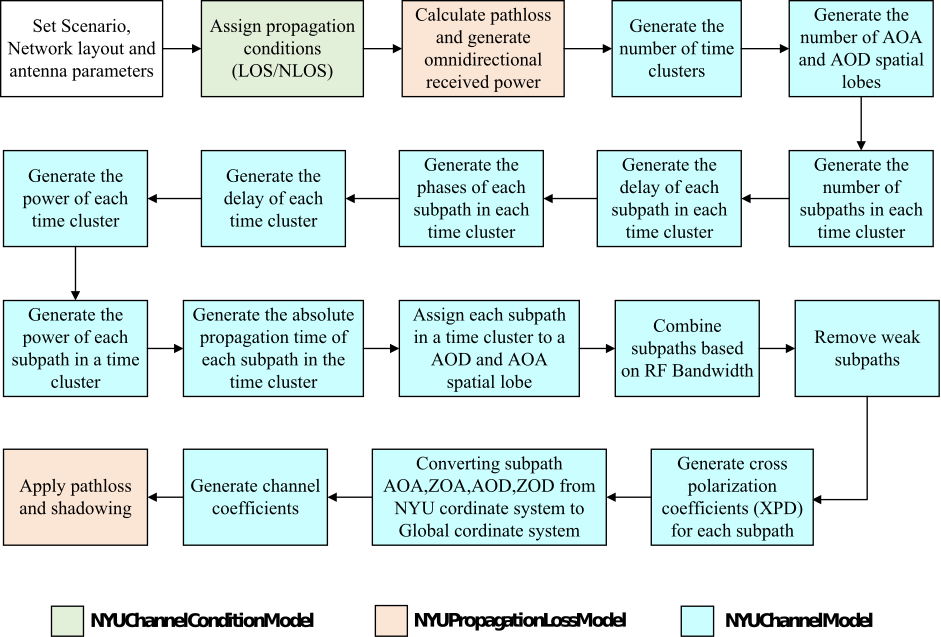}}
\caption{Channel Generation Procedure for NYUSIM; The Colors Indicate the Class where the Steps Are Accomplished}
\label{fig:chanGenProcedure}
\end{figure*}
The first step in generating a channel model for a UE is determining whether the channel condition is LOS/NLOS. We extend the \texttt{Channel\-Condition} class developed in \cite{zugno2020implementation} with five different classes, namely, \texttt{NYU\-Rma\-Channel\-Condition\-Model}, \texttt{NYU\-Uma\-Channel\-Condition\-Model}, \texttt{NYU\-Umi\-Channel\-Condition\-Model}, \texttt{NYU\-InH\-Channel\-Condition\-Model}, and \texttt{NYU\-InF\-Channel\-Condition\-Mod\-el}, each handling a different scenario for NYUSIM. All the newly introduced NYUSIM classes mentioned above derive from the same base class, called \texttt{NYU\-Channel\-Condi\-tion}, which extends the \texttt{Cha\-nnel\-Condition\-Model} interface. The NYUSIM LOS probability models for outdoor scenarios were developed based on radio propagation measurements conducted at 28 and 73 GHz in New York City \cite{7070688}. The Tx-Rx locations were selected from measurement data, and ray tracing was employed to determine whether the path between the Tx and Rx is in LOS/NLOS \cite{7070688}. For the UMi and UMa scenarios, 3GPP SCM had a higher probability of predicting LOS channel conditions at larger distances (several hundred meters) than NYUSIM \cite{rappaport2017investigation}. However, in a real-world UMi and UMa scenario, getting LOS propagation over several hundred meters is very challenging \cite{7070688,rappaport2017investigation}. Thus, the UMi and UMa, LOS probability models, implemented in \texttt{NYU\-Umi\-Channel\-Condition\-Model} and \texttt{NYU\-Uma\-Chan\-nel\-Conditi\-on\-Model}, use the NYU (squared) model based on Tables 1 and 2 in \cite{Rap2017itap,7070688} respectively and is given in  \eqref{LosProbUmi} and \eqref{LosProbUma}. 

\begin{align}
    \label{LosProbUmi}
    \mathrm{P_{LOS}}(d_{2D}) =  (\mathrm{min}(d_{1}/d_{2D},1)&(1-\mathrm{exp}(-d_{2D}/d_{2})) \nonumber\\&+ \mathrm{exp}(-d_{2D}/d_{2}))^2,
\end{align}
\begin{align}
    \label{LosProbUma}
    \mathrm{P_{LOS}}(d_{2D}) =  &((\mathrm{min}(d_{1}/d_{2D},1)(1-\mathrm{exp}(-d_{2D}/d_{2})) \nonumber \\& + \mathrm{exp}(-d_{2D}/d_{2}))(1 + C(d_{2D},h_{UE})))^2.
\end{align}
In \eqref{LosProbUmi} $d_{1}=$ 22m, $d_{2}$ = 100m, in \eqref{LosProbUma} $d_{1}=$ 20m, $d_{2}$ = 160m, $C$ is defined in Table 7.4.2 in \cite{3GPPTR} and $h_{UE}$ is the height of the UE in meters. For both \eqref{LosProbUmi} and \eqref{LosProbUma} the values of $d_{1}$, $d_{2}$ are based on measurement data \cite{7070688} and $d_{2D}$ is the 2D separation distance between the Tx-Rx in meters. However, LOS probability models for InH, RMa, and InF scenarios do not exist in NYUSIM. For the InH LOS probability models in \texttt{NYU\-InH\-Channel\-Condition\-Model}, we use the 5GCM model from Table 3 \cite{Rap2017itap} because the NYU (squared) model for UMi and UMa scenarios is similar to that of 5GCM LOS probability models and 5GCM using a similar curve fitting approach as NYUSIM to develop LOS probability models. In contrast, for the RMa scenario in \texttt{NYU\-Rma\-Channel\-Condition\-Model}, we use the LOS probability models given by 3GPP in Table 7.4.2 \cite{3GPPTR} since 5GCM doesn't support LOS probability models for RMa scenario. The \texttt{NYUInFChannelConditionModel} takes an approach of using the average of the LOS probability models for InF-SL, InF-DL, InF-SH and InF-DH scenarios, which are provided by 3GPP in Table 7.4.2 \cite{3GPPTR}. This is done because 5GCM doesnt support LOS probability models for InF scenario and NYUSIM does not classify factory environments into sub-scenarios like InF-SL, InF-DL, InF-SH and InF-DH scenarios based on size, clutter density, and clutter height, as specified in 3GPP \cite{3GPPTR}. Therefore, the average of the LOS probability models is used as a general representation of InF scenarios in the \texttt{NYUInFChannelConditionModel}. The implementation of all the methods mentioned in this section is present in the file \texttt{nyu-cha\-nne\-l-co\-ndi\-tio\-n-mo\-del.cc}.

\subsection{Large-scale Path Loss Models}
\label{pathLossModels}
The received power as a function of the distance from the transmitter is characterized using large-scale path loss models \cite{rappaport2010wireless}. The large-scale model parameters (path loss and shadowing) refer to the characteristics of the wireless channel that vary slowly over time and space \cite{rappaport2010wireless}. To calculate the path loss for all scenarios except RMa scenario, NYUSIM uses the close-in (CI) free space reference distance path loss model with a 1 m reference distance, which was found to have superior accuracy and less sensitivity to error from measurement campaigns in many different scenarios \cite{sun2016investigation}. The expression for the CI path loss model \cite{rappaport2015wideband,sun2016investigation,maccartney2015indoor,maccartney2017study} with additional attenuation terms due to atmospheric attenuation (AT) \cite{liebe1993propagation}, outdoor to indoor loss (O2I) loss \cite{ju2019millimeter} and foliage loss (FL) \cite{7247347,hp2023icc} in our ns-3 implementation of NYUSIM is given in \eqref{ciPathLossModel}:
\begin{align}
    \label{ciPathLossModel}
    \mathrm{PL^{CI}}(f,d) [\mathrm{dB}] = \; & \mathrm{FSPL} (f,1 \; \mathrm{m}) [\mathrm{dB}] +  10 \eta \log_{10}(d) \nonumber \\
    &+ \mathrm{AT}[\mathrm{dB}] + \mathrm{O2I} [\mathrm{dB}] \\ \nonumber &+ \mathrm{FL} [\mathrm{dB}] + \chi_{\sigma}^{CI} [\mathrm{dB}],
\end{align}
where $f$ is the frequency in GHz, $d$ denotes the 2D Tx-Rx separation distance, and $d \geq 1$ m, $\mathrm{FSPL}(f,1 \; \mathrm{m})$ represents the free space path loss at a Tx-Rx separation of 1 m at carrier frequency $f$, $\eta$ denotes the path loss exponent (PLE). O2I loss is caused due to the propagation of the signal from the outdoor to the indoor environment. O2I loss is implemented using the high and low loss parabolic models for building penetration loss \cite{haneda20165g,5GCMChannelModel}. $\chi_{\sigma}^{CI}$ is a zero-mean Gaussian random variable with $\sigma$ as the standard deviation in dB to represent shadow fading about the distant-dependent mean path loss value.
\par For the RMa scenario to compute the large-scale path loss we use the CIH path loss model (CI model with a height depend PLE). Mathematically the expression for CIH path loss model in LOS and NLOS channel condition for RMa scenario can be expressed as \eqref{cihPathLossModelLos} and \eqref{cihPathLossModelNlos}, respectively,

 \begin{align}
     \label{cihPathLossModelLos}
     \mathrm{PL^{CIH}}(f,d) [\mathrm{dB}] = \ & \mathrm{FSPL} (f,1 m) [\mathrm{dB}] + \nonumber 23.1\left(1-0.03\left(\frac{h_{BS}-35}{35}\right)\right)\nonumber \\& \times \log_{10}(d) + \chi_{\sigma}^{CI} [\mathrm{dB}],
 \end{align}
 
  \begin{align}
     \label{cihPathLossModelNlos}
     \mathrm{PL^{CIH}}(f,d) [\mathrm{dB}] = \ & \mathrm{FSPL} (f,1 m) [\mathrm{dB}] + \nonumber 30.7\left(1-0.049\left(\frac{h_{BS}-35}{35}\right)\right) \nonumber \\& \times \log_{10}(d) + \chi_{\sigma}^{CI} [\mathrm{dB}],
 \end{align}
where $h_{BS}$ is the height of the base station in meters and other terms are defined similarly to \eqref{ciPathLossModel}.
\par To implement the large-scale models, we developed the base class \texttt{NYU\-Propagation\-Loss\-Model}, which extends the \texttt{Propagation\-Loss\-Model} interface. Then, we extended \texttt{NYU\-Pro\-pag\-ati\-on\-Loss\-Model} by developing four sub classes, i.e., \texttt{NYU\-Rma\-Pro\-pag\-at\-ion\-Loss\-Model}, \texttt{NYU\-Uma\-Propagation\-Loss\-Model}, \texttt{NYU\-Umi\-Pro\-pag\-at\-ion\-Loss\-Model}, \\\texttt{NYU\-InH\-Propagation\-Loss\-Model}, and \texttt{NYU\-InF\-Propagation\-Loss\-Mo\-del}, which compute the path loss for RMa, UMa, UMi, InH, and InF scenarios, respectively. All the implementation of all the classes mentioned above are present in the file \texttt{nyu\--pr\-opa\-gat\-ion\--lo\-ss\--mod\-el.cc}. We pair the \texttt{NYU\-Propagation\-Loss\-Model} with the channel condition model through the \texttt{Channel\-Condition\-Model} interface to model the interdependence of the path loss to the LOS/NLOS conditions. Specifically, the method \texttt{DoCalcRxPower} computes the total path loss according to \eqref{ciPathLossModel} via calls to other methods. First, \texttt{DoCalcRxPower} uses either \texttt{GetLossLos} or \texttt{GetLossNlos} based on the scenario and depending on whether the channel is in LOS or NLOS to calculate the terms in \eqref{ciPathLossModel} or \eqref{cihPathLossModelLos} or \eqref{cihPathLossModelNlos}. The value of $\eta$ is based on the channel condition (LOS/NLOS), the scenario (UMi, UMa, InH, and InF), and the operating frequency whereas for RMa scenario $h_{BS}$ is taken into consideration. If any of the additional attenuation's is enabled, \texttt{Do\-Calc\-Rx\-Power} invokes the corresponding methods \texttt{Get\-O2I\-Path\-Loss}, \texttt{Get\-Foliage\-Loss} and \texttt{Get\-Atmospheric\-Attenuation} to compute O2I [dB], FL [dB], and AT [dB], respectively. \par We use the method \texttt{Get\-Shadowing} in \texttt{Do\-Calc\-Rx\-Power} if shadowing loss is enabled to apply the shadowing model. Two other functions, namely \texttt{Get\-Shadowing\-Std} and \texttt{Get\-Shadowing\-Correlation\-Distance} are used by \texttt{Get\-Shadowing} to retrieve the standard deviation of the shadow fading component ($\chi^{CI}_{\sigma}$) and the shadowing correlation distance. An exponential autocorrelation function is used for correlating adjacent shadow fading values. The exponential autocorrelation depends on the spatial separation between the two nodes (gNB and UE) as described in \cite{3GPPTR}.\par As NYUSIM is based on measurements at few different frequency bands (28, 38, 73, and 140 GHz) and all scenarios except InF (InF measurement is only done at 142 GHz) are only measured for 28 GHz and 140 GHz, it is imperative to interpolate values for any parameter such as PLE and $\chi^{CI}_{\sigma}$ for any arbitrary frequency between 28 GHz and 140 GHz using the expression in \eqref{linearInterpolation}:
\begin{align}
    \label{linearInterpolation}
    \resizebox{.9\hsize}{!}{$
    	p(f)=\left\{\begin{matrix}
		p(28) & ,f \leqslant 28,\\ 
		\frac{p(140)-p(28)}{140-28}f+\frac{5p(28)-p(140)}{4} & , 28 < f < 140,\\ 
		p(140) & , f \geqslant 140,
	\end{matrix}\right.
	$}
\end{align}
where $p(f)$ denotes the PLE or $\chi^{CI}_{\sigma}$ value at frequency $f$ in GHz, $p(28)$ and $p(140)$ represent the PLE or $\chi^{CI}_{\sigma}$ value at 28 GHz and 140 GHz, respectively. The values of PLE, and $\chi^{CI}_{\sigma}$ at 28 GHz and 140 GHz for different scenarios and channel conditions are specified in Tables \ref{tab:28GHz values} and \ref{tab:142GHz values} and were empirically determined by curve fitting using measurement data at 28 GHz and 140 GHz \cite{rappaport2013millimeter,maccartney2015indoor,maccartney2016millimeter,maccartney2017rural,ju20203,9558848}. For the InF scenario we use the values in Table \ref{tab:142GHz values} across the entire frequency range of 0.5 to 150 GHz.
Furthermore, we use the spatial correlation distances based on LOS/NLOS channel condition as specified in 3GPP TR 38.901 \cite{3GPPTR} in all 3GPP listed scenarios to compute the correlated adjacent shadow fading values.

\subsection{Small-scale Multipath Models and Channel Generation Procedure for NYUSIM}
\label{sec:changen}
Small-scale parameters refer to the characteristics of the wireless channel that vary rapidly over time and space \cite{rappaport2010wireless}. The small-scale multipath (MP) models are needed to accurately model the variations in the wireless channel that occur over short distances and periods of time and help in statistically reproducing the CIR by capturing the MP components (MPCs) in the time-varying wireless channel. NYUSIM uses the Time Cluster Spatial Lobe (TCSL) approach described in \cite{samimi20163}, whereas 3GPP SCM uses a joint delay angle probability density function \cite{3GPPTR} to generate the temporal and spatial characteristics of the channel. The TCSL approach is borne out of
extensive measurement campaigns by NYU WIRELESS from
2011-2022. A Time Cluster (TC) is comprised of MPCs traveling closely in time
(tens or hundreds of nanoseconds) \cite{samimi20163}, and Spatial Lobe (SL) denotes the direction of arrival/departure of most energy over the azimuth or elevation plane \cite{samimi20163}. We combine the channel condition and path loss models with the small-scale characteristics described in Figure \ref{fig:chanGenProcedure} to generate the channel using NYUSIM. A detailed description of each step is presented below:\\
\begin{itemize}
    \item \textit{\textbf{Step 1}}: Based on the scenario specified in the simulation script, the channel condition is computed (LOS/NLOS) using the LOS probability models described in Section \ref{losModels}.
    \item  \textit{\textbf{Step 2}}: Once the channel condition is set, the large-scale path loss is computed using the CI or CIH path loss model and additional attenuation terms as outlined in Section \ref{pathLossModels}.
    \item \textit{\textbf{Step 3}}: \texttt{Get\-Number\-Of\-Time\-Clusters} generates the number of TCs based on distributions in Table \ref{Distributions} - Step 1.
    \item \textit{\textbf{Step 4}}: \texttt{Get\-Number\-Of\-Aoa\-Spatial\-Lobes} and \texttt{Get\-Number\-Of\-Aod\-Spatial\-Lobes} generates the number of the Angle of Arrival (AOA) and Angle of Departure (AOD) SLs, respectively, based on distributions in Table \ref{Distributions} - Step 8. For example, 1 AOA SL lobe indicates that there is one SL in azimuth and elevation.
    \item \textit{\textbf{Step 5}}: The number of MPs in each TC is generated in \texttt{Get\-Number\-Of\-Subpaths\-In\-Time\-Cluster} using the distribution specified in Table \ref{Distributions} - Step 2.
    \item \textit{\textbf{Step 6}}: The Time of Arrival (TOA) of each MP in a TC is computed in \texttt{Get\-Intra\-Cluster\-Delays} using the distribution specified in Table \ref{Distributions} - Step 4.
    \item \textit{\textbf{Step 7}}: The phase of MP for each polarization (V-V, V-H, H-V, and H-H) is described using a uniform random variable in (0,2$\pi$) in the method \texttt{Get\-Subpath\-Phases}. 
    \item \textit{\textbf{Step 8}}: \texttt{Get\-Cluster\-Excess\-Time\-Delays} computes the time delay of each TC based on the distribution shown in Table \ref{Distributions} - Step 3.
    \item \textit{\textbf{Step 9}}: To generate the total power of each TC in \texttt{Get\-Cluster\-Powers}, we use the distribution mentioned in Table \ref{Distributions} - Step 5.
    \item \textit{\textbf{Step 10}}: \texttt{Get\-Subpath\-Powers} uses the distribution specified in Table \ref{Distributions} -  Step 6 to calculate the power of each MP.
    \item \textit{\textbf{Step 11}}: The absolute propagation delay of the MP is computed in \texttt{Get\-Absolute\-Propagation\-Times} based on the actual distance between the Tx and Rx. The total delay of each MP is the sum of the absolute delay of the MP and the delay of the MP computed in Section \ref{sec:changen} - Step 6.
    \item \textit{\textbf{Step 12}}: In \texttt{Get\-Subpath\-Mapping\-And\-Angles}, we compute the azimuth AOD, the zenith angle of departure (ZOD), the azimuth AOA and zenith angle of arrival (ZOA) for each MP based on the distribution specified in Table \ref{Distributions} - Step 9, Step 10 and Step 11.
    \item \textit{\textbf{Step 13}}: In wideband systems, more MPs can be resolved in time, whereas in narrowband systems, only a few MPs can be resolved. Thus, \texttt{Get\-BW\-Adjustedted\-Power\-Spectrum} vectorially combines MPs based on the bandwidth of operation. Secondly, once the resolvable MPs are obtained, in LOS conditions, the azimuth AOA and AOD, the ZOA and ZOD of the first MP must be aligned with each other \cite{sun2015synthesizing}; hence, in LOS conditions, we call the method \texttt{Get\-Los\-Aligned\-Power\-Spectrum} to perform the geometrical alignment of the first MP component in azimuth and elevation.
    \item \textit{\textbf{Step 14}}: The receiver can detect powers only above a certain threshold value ($P_{RX,Th}$). Thus, in the method \texttt{Get\-Valid\-Subapths}, we remove MPs with powers less than maximum MP power - $P_{RX,Th}$. The value of $P_{RX,Th}$ is calculated in \texttt{Dynamic\-Range}. The dynamic range of the NYU channel sounder is determined based on the maximum measurable omnidirectional path loss at 28 GHz in LOS and is independent of scenario.
    \item \textit{\textbf{Step 15}}: \texttt{Get\-Xpd\-Per\-Subpath} generates the cross-polarization (XPD) coefficients values for H-H, V-H, and H-V with respect to (w.r.t) V-V polarization for each MP.
    \item \textit{\textbf{Step 16}}: We use the 3GPP antenna model \cite{rebato2018multi}, which uses the global coordinate system (GCS) described in 3GPP TR 38.901 \cite{3GPPTR} while generating the channel matrix for NYUSIM. In GCS, the azimuth angle is measured w.r.t the x-axis, and the elevation angle is measured w.r.t the z-axis. However, in NYUSIM, azimuth and elevation angle is measured w.r.t the y-axis. Thus, we convert the generated AOA, AOD, ZOA, and ZOA for each MP to the GCS using the method \texttt{NYU\-Cordinate\-System\-To\-Global\-Cordinate\-System}.
    \item \textit{\textbf{Step 17}}: Finally, in \texttt{Get\-New\-Channel} we compute the MIMO channel matrix for NYUSIM. The details are presented in Section \ref{sec:channelMatrix}.
\end{itemize}
\begin{figure*}[h]
  \begin{subfigure}{0.195\linewidth}
    \includegraphics[width=\linewidth]{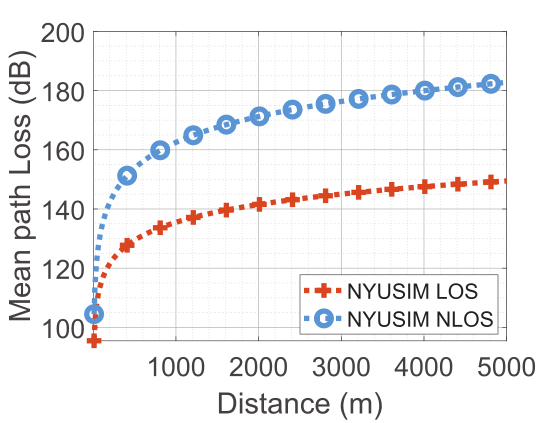}
    \caption{Mean path loss for UMi}
  \end{subfigure}
  \begin{subfigure}{0.195\linewidth}
    \includegraphics[width=\linewidth]{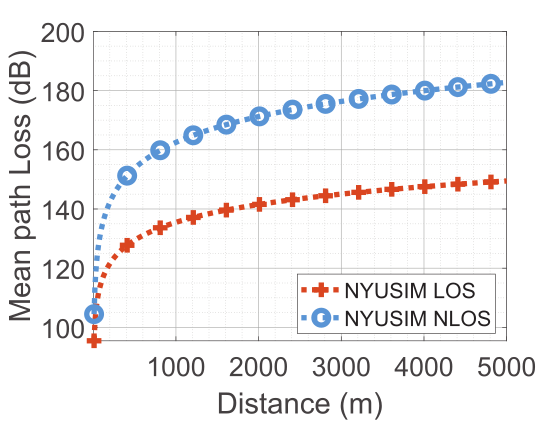}
    \caption{Mean path loss for UMa}
  \end{subfigure}
  \begin{subfigure}{0.195\linewidth}
    \includegraphics[width=\linewidth]{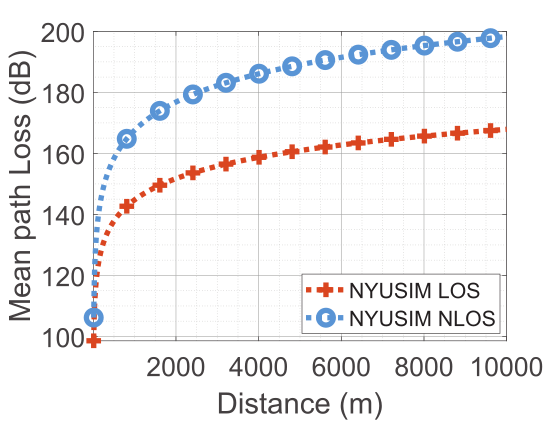}
    \caption{Mean path loss for RMa}
  \end{subfigure}
  \begin{subfigure}{0.195\linewidth}
    \includegraphics[width=\linewidth]{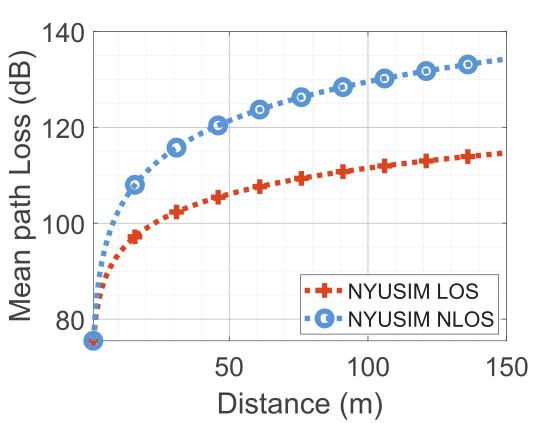}
    \caption{Mean path loss for InH}
  \end{subfigure}
  \begin{subfigure}{0.195\linewidth}
    \includegraphics[width=\linewidth]{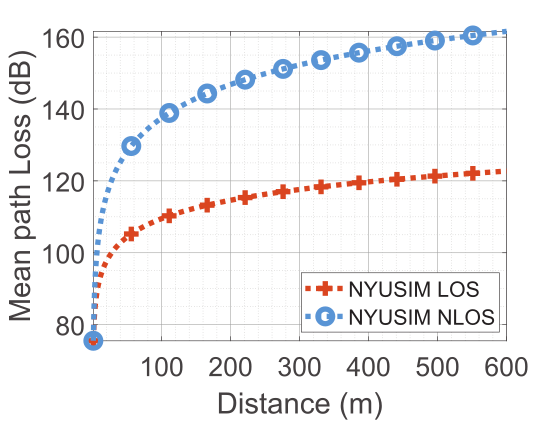}
    \caption{Mean path loss for InF}
  \end{subfigure}
  \caption{Large-scale Mean Path Loss (dB) vs Distance (m) at 142 GHz using NYUSIM for all 3GPP-listed Scenarios in LOS/NLOS}
   \label{fig:PathLoss}
\end{figure*}
\vspace{-0.1in}
\par The statistical distributions and parameter values used for generating the small-scale MP models for NYUSIM in ns-3 are based on MATLAB implementation of \textit{NYUSIM}, which has been used by researchers worldwide since 2016. The distributions in Table \ref{tab:distributions} are characterized using the parameter values listed in Tables \ref{tab:28GHz values} and \ref{tab:142GHz values} at 28 GHz and 140 GHz, respectively. All the distributions (Table \ref{tab:distributions}), parameter values (Tables \ref{tab:28GHz values} and \ref{tab:142GHz values}), and the methods (steps 1-17) are implemented in the file \texttt{nyu-\-channel-\-model.cc}. To find the small-scale MP model parameters for any specific frequency, we use \eqref{linearInterpolation}. 

\renewcommand{\arraystretch}{1}
\renewcommand{\arraystretch}{1.4}
\begin{table*}[h!]
  \centering
  \caption{Channel Parameters for NYUSIM at 28 GHz in all 3GPP-listed Scenarios (UMi, UMa, RMa, InH and InF)}
  \resizebox{1.7\columnwidth}{!}{%
    \begin{tabular}{|O|O|N|N|N|N|N|N|N|N|cc|}
    \hline
    \multicolumn{2}{|c|}{\multirow{2}[3]{4cm}{\centering \textbf{Channel Parameter}}} & \multicolumn{2}{c|}{\textbf{UMi}} & \multicolumn{2}{c|}{\textbf{UMa}} & \multicolumn{2}{c|}{\textbf{RMa}} & \multicolumn{2}{c|}{\textbf{InH}} & \multicolumn{2}{c|}{\textbf{InF}} \\
\cline{3-12}    \multicolumn{2}{|c|}{} & \multicolumn{1}{c|}{LOS} & \multicolumn{1}{c|}{NLOS} & \multicolumn{1}{c|}{LOS} & \multicolumn{1}{c|}{NLOS} & \multicolumn{1}{c|}{LOS} & \multicolumn{1}{c|}{NLOS} & \multicolumn{1}{c|}{LOS} & \multicolumn{1}{c|}{NLOS} & \multicolumn{1}{c|}{LOS} & NLOS \\
    \hline
    Path loss & $n$     & 2     & 3.2   & 2     & 2.9   & NA   & NA   & 1.2   & 2.7   & \multicolumn{2}{c|}{\multirow{14}[25]{1cm}{Data not available at 28 GHz}} \\
\cline{1-10}    Shadow fading & $\chi_{\sigma}^{CI}$    & 4     & 7     & 4     & 7     & 1.7   & 6.7   & 3     & 9.8   & \multicolumn{2}{c|}{} \\
\cline{1-10}    Number of time clusters & $N_c, \lambda_c$ & $N_c$ = 6 & $N_c$ = 6 & $N_c$ = 6 & $N_c$ = 6 & $N_c$ = 1 & $N_c$ = 1 & $\lambda_c$ = 3.6 & $\lambda_c$ = 5.1 & \multicolumn{2}{c|}{} \\
\cline{1-10}    \multirow{2}[1]{2cm}{\centering Number of cluster subpaths} & \multirow{2}[2]{*}{$M_s, \mu_s,\beta_s$} & \multirow{2}[2]{*}{$M_s$ = 30} & \multirow{2}[2]{*}{$M_s$ = 30} & \multirow{2}[2]{*}{$M_s$ = 30} & \multirow{2}[2]{*}{$M_s$ = 30} & \multirow{2}[2]{*}{$M_s$ = 2} & \multirow{2}[2]{*}{$M_s$ = 2} & $\beta_s$ = 0.7 & $\beta_s$ = 0.7 & \multicolumn{2}{c|}{} \\
          &       &       &       &       &       &       &       & $\mu_s$ = 3.7 & $\mu_s$ = 5.3 & \multicolumn{2}{c|}{} \\
\cline{1-10}    \centering Cluster excess delay & $\mu_\tau$ [ns] & 123   & 83    & 123   & 83    & 123   & 83    & 17.3  & 10.9  & \multicolumn{2}{c|}{} \\
\cline{1-10}    \centering Intra-cluster subpath excess delay & $X, \mu_\rho$ [ns] & $X$ = 0.2 & $X$ = 0.5 & $X$ = 0.2 & $X$ = 0.5 & $X$ = 0.2 & $X$ = 0.5 & $\mu_\rho$ = 3.4 & $\mu_\rho$ = 22.7 & \multicolumn{2}{c|}{} \\
\cline{1-10}    \centering Cluster power & $\Gamma$ [ns], $\sigma_Z$ [dB] & 25.9, 1 & 51, 3  & 25.9, 1 & 51, 3  & 25.9, 1 & 51, 3  & 20.7, 10 & 23.6, 10 & \multicolumn{2}{c|}{} \\
\cline{1-10}    \centering Subpath power & $\gamma$ [ns], $\sigma_U$ [dB] & 16.9, 6 & 15.5, 6 & 16.9, 6 & 15.5, 6 & 16.9, 6 & 15.5, 6 & 2, 5 & 9.2, 6 & \multicolumn{2}{c|}{} \\
\cline{1-10}    \centering Number of spatial lobes & $\lambda_{l,AOD}, \lambda_{l,AOA}$ & 1.9, 1.8 & 1.5, 2.1 & 1.9, 1.8 & 1.5, 2.1 & 1, 1   & 1, 1   & 3, 3   & 3, 3   & \multicolumn{2}{c|}{} \\
\cline{1-10}    \multirow{2}[0]{2cm}{\centering Mean lobe elevation angle [$^o$]} & $\mu_{l,ZOD}, \sigma_{l,ZOD}$ & -12.6, 5.9 & -4.9, 4.5 & -12.6, 5.9 & -4.9, 4.5 & -12.6, 5.9 & -4.9, 4.5 & -7.3, 3.8 & -5.5, 2.9 & \multicolumn{2}{c|}{} \\
\cline{2-10}          & $\mu_{l,ZOA}, \sigma_{l,ZOA}$ & 10.8, 5.3 & 3.6, 4.8 & 10.8, 5.3 & 3.6, 4.8 & 10.8, 5.3 & 3.6, 4.8 & 7.4, 3.8 & 5.5, 2.9 & \multicolumn{2}{c|}{} \\
\cline{1-10}    \multirow{2}[3]{2cm}{\centering Subpath angular offset [$^o$]} & $\sigma_{\phi,AOD},\sigma_{\theta,ZOD}$ & 8.5, 2.5 & 11, 3  & 8.5, 2.5 & 11, 3  & 8.5, 2.5 & 11, 3  & 20.6, 15.7 & 27.1, 16.2 & \multicolumn{2}{c|}{} \\
\cline{2-10}          & $\sigma_{\phi,AOA}, \sigma_{\theta,ZOA}$ & 10.5, 11.5 & 7.5, 6 & 10.5, 11.5 & 7.5, 6 & 10.5, 11.5 & 7.5, 6 & 17.7, 14.4 & 20.3, 15.0 & \multicolumn{2}{c|}{} \\
\hline
    \end{tabular}%
    }
  \label{tab:28GHz values}%
\end{table*}%
\renewcommand{\arraystretch}{1}

\renewcommand{\arraystretch}{1.45}
\begin{table*}[h!]
  \centering
  \caption{Channel Parameters for NYUSIM at 140 GHz in all 3GPP-listed Scenarios (UMi, UMa, RMa, InH and InF)}
  \resizebox{1.7\columnwidth}{!}{%
    \begin{tabular}{|P|Q|N|N|N|N|N|N|N|N|N|N|}
\cline{1-12}    \multicolumn{2}{|c|}{\multirow{2}[1]{3.5cm}{\centering \textbf{Channel Parameter}}} & \multicolumn{2}{c|}{\textbf{UMi}} & \multicolumn{2}{c|}{\textbf{UMa}} & \multicolumn{2}{c|}{\textbf{RMa}} & \multicolumn{2}{c|}{\textbf{InH}} & \multicolumn{2}{c|}{\textbf{InF}} \\
\cline{3-12}    \multicolumn{2}{|c|}{} & LOS   & NLOS  & LOS   & NLOS  & LOS   & NLOS  & LOS   & NLOS  & LOS   & NLOS \\
    \hline
    Path loss &  $n$     & 2     & 2.9   & 2     & 2.9   & NA   & NA   & 1.8   & 2.7   & 1.7   & 3.1 \\
    \hline
    Shadow fading & $\chi_{\sigma}^{CI}$      & 2.6   & 8.2   & 2.6   & 8.2   & 1.7   & 6.7   & 2.9   & 6.6   & 3     & 7 \\
    \hline
    Number of time clusters &  $N_c, \lambda_c$      & $N_c$ = 5 & $N_c$ = 3     & $N_c$ = 5 & $N_c$ = 3     & $N_c$ = 1 & $N_c$ = 1 & $\lambda_c$ = 0.9   & $\lambda_c$ = 1.8   & $\lambda_c$ = 2.4   & $\lambda_c$ = 2 \\
    \hline
    \multirow{4}[2]{1.7cm}{\centering Number of cluster subpaths} & \multirow{4}[2]{1.7cm}{\centering $\mu_s, M_s,$ $\beta_s$} & \multirow{4}[2]{*}{$\mu_s$ = 1.8} & \multirow{4}[2]{*}{$\mu_s$ = 3} & \multirow{4}[2]{*}{$\mu_s$ = 1.8} & \multirow{4}[2]{*}{$\mu_s$ = 3} & \multirow{4}[2]{*}{$M_s$ = 2} & \multirow{4}[2]{*}{$M_s$ = 2} & \multirow{2}[1]{*}{$\beta_s$ = 1} & \multirow{2}[1]{*}{$\beta_s$ = 1} & \multirow{2}[1]{*}{$\beta_s$ = 1} & \multirow{2}[1]{*}{$\beta_s$ = 1} \\
          &       &       &       &       &       &       &       &       &       &  &  \\
          &       &       &       &       &       &       &       & \multirow{2}[1]{*}{$\mu_s$ = 1.4} & \multirow{2}[1]{*}{$\mu_s$ = 1.2} & \multirow{2}[1]{*}{$\mu_s$ = 2.6} &  \multirow{2}[1]{*}{$\mu_s$ = 7}\\
          &       &       &       &       &       &       &       &       &       &  &  \\
    \hline
    
    \multirow{2}[1]{1.7cm}{ \centering Cluster excess delay [ns]} & \multirow{2}[1]{1.7cm}{\centering $\mu_\tau$, $\alpha_\tau$, $\beta_\tau$} & \multirow{2}[1]{*}{$\mu_\tau$ = 80} & \multirow{2}[1]{*}{$\mu_\tau$ = 58} & \multirow{2}[1]{*}{$\mu_\tau$ = 80} & \multirow{2}[1]{*}{$\mu_\tau$ = 58} & \multirow{2}[1]{*}{$\mu_\tau$ = 80} & \multirow{2}[1]{*}{$\mu_\tau$ = 58} & \multirow{2}[1]{*}{$\mu_\tau$ = 14.6} & \multirow{2}[1]{*}{$\mu_\tau$ = 21} & $\alpha_\tau$ = 0.7 & $\alpha_\tau$ = 0.8 \\
          &       &       &       &       &       &       &       &       &       & $\beta_\tau$ = 26.9 & $\beta_\tau$ = 13.9 \\
    \hline
    \multirow{2}[2]{1.7cm}{\centering Intra-cluster subpath excess delay [ns] } & \multirow{2}[2]{1.7cm}{\centering $\mu_\rho, \alpha_\rho,\beta_\rho $} & \multirow{2}[2]{*}{$\mu_\rho$ = 30} & \multirow{2}[2]{*}{$\mu_\rho$ = 33} & \multirow{2}[2]{*}{$\mu_\rho$ = 30} & \multirow{2}[2]{*}{$\mu_\rho$ = 33} & \multirow{2}[2]{*}{$\mu_\rho$ = 30} & \multirow{2}[2]{*}{$\mu_\rho$ = 33} & \multirow{2}[2]{*}{$\mu_\rho$ = 1.1} & \multirow{2}[2]{*}{$\mu_\rho$ = 2.7} & $\alpha_\rho$ = 1.2 & $\alpha_\rho$ = 1.6 \\
          &       &       &       &       &       &       &       &       &       & $\beta_\rho$ = 16.3 & $\beta_\rho$ = 9 \\
    \hline
    Cluster power &   $\Gamma$ [ns], $\sigma_Z$ [dB]     & 40, 5.34 & 49, 4.68 & 40, 5.34 & 49, 4.68 & 40, 5.34 & 49, 4.68 & 18.2, 9 & 16.1, 10 & 16.2,10 & 18.7,6 \\
    \hline
    Subpath power &   $\gamma$ [ns], $\sigma_U$ [dB]    & 20, 3.48 & 37, 3.62 & 20, 3.48 & 37, 3.62 & 20, 3.48 & 37, 3.62 & 2, 5  & 2.4, 6 & 4.7, 13 & 7.3, 11 \\
    \hline
    Number of spatial lobes &  \centering $\lambda_{l,AOD},\lambda_{l,AOA}$     & 1.4, 1.2 & 1.3, 2.1 & 1.4, 1.2 & 1.3, 2.1 & 1, 1 & 1, 1 & 2, 2  & 3, 2   & 1.8, 1.9 & 1.8, 2.5 \\
    \hline
    \multirow{2}[1]{1.7cm}{\centering Mean lobe elevation angle [$^o$]} & $\mu_{l,ZOD},\sigma_{l,ZOD}$     & -3.2, 1.2 & -1.6, 0.5 & -3.2, 1.2 & -1.6, 0.5 & -3.2, 1.2 & -1.6, 0.5 & -6.8, 4.9 & -2.5, 2.7 & -4, 4.3 & -3, 3.5 \\
\cline{2-12}          &   $\mu_{l,ZOA},\sigma_{l,ZOA}$   & 2, 2.9 & 1.6, 2 & 2, 2.9 & 1.6, 2 & 2, 2.9 & 1.6, 2 & 7.4, 4.5 & 4.8, 2.8 & 4, 4.3 & 3, 3.5 \\
\hline
    \multirow{2}[1]{1.7cm}{\centering Subpath angular offset [$^o$]} &$\sigma_{\phi, AOD}$, $\sigma_{\theta, ZOD}$     & 4.3, 0.1 & 5, 2.3 & 4.3, 0.1 & 5, 2.3 & 4.3, 0.1 & 5, 2.3 & 4.8, 4.3 & 4.8, 2.8 & 6.7, 3 & 9.3, 4.5 \\
\cline{2-12}          &$\sigma_{\phi, AOA}$, $\sigma_{\theta, ZOA}$   & 7.3, 3.2 & 7.5, 0 & 7.3, 3.2 & 7.5, 0 & 7.3, 3.2 & 7.5, 0 & 4.7, 4.4 & 6.6, 4.5 & 11.7, 2.3 & 14.1, 3.2 \\
    \hline
    \end{tabular}%
    }
  \label{tab:142GHz values}%
\end{table*}%
\renewcommand{\arraystretch}{1}

\vspace{-0.1in}
\subsection{Channel Matrix for NYUSIM}
\label{sec:channelMatrix}
All the paths between the Tx and Rx antennas are captured using a wireless channel matrix \textbf{H}. 
It provides a mathematical representation of the wireless channel that can be used to analyze the performance of the wireless communication system and optimize its design. A \textbf{H} similar to 3GPP TR 38.901 \cite{rebato2018multi,3GPPTR} is implemented for NYUSIM in ns-3 and is expressed in \eqref{nyuChannelMatrix}
\begin{align}
    \label{nyuChannelMatrix}
    H(\tau,t) = \sum_{m = 1}^{M} H_{u,s,m}(t) \delta(\tau - \tau_{m}),
\end{align}
where $M$ denotes the total number of MPCs, $m$ represents the $m^{th}$ MP, and $\tau_{m}$ is the delay of the $m^{th}$ MP. The channel coefficients $H_{u,s,m}(t)$ at time $t$ is computed using \eqref{nyuChannelCoefficient}
\begin{align}
    \label{nyuChannelCoefficient}
    &H_{u,s,m}(t) =  \alpha_{m}
    \begin{bmatrix}
    F_{rx,u,\theta}(\theta_{m,ZOA},\phi_{m,AOA}) \\ \nonumber
    F_{rx,u,\phi}(\theta_{m,ZOA},\phi_{m,AOA})
    \end{bmatrix}^\top \\  \nonumber
    &\begin{bmatrix}
    exp(j\phi_{m}^{\theta\theta}) & \sqrt{\frac{1}{K_{m_{\theta,\phi}}}} \exp(j \phi_{m}^{\theta\phi})\\
     \sqrt{\frac{1}{K_{m_{\phi,\theta}}}}exp(j \phi_{m}^{\phi\theta}) & \sqrt{\frac{1}{K_{m_{\phi,\phi}}}} \exp(j \phi_{m}^{\phi\phi})
     \end{bmatrix}\\
    &\begin{bmatrix}
     F_{tx,s,\theta}(\theta_{m,ZOD},\phi_{m,AOD}) \\
     F_{tx,s,\phi}(\theta_{m,ZOD},\phi_{m,AOD})
     \end{bmatrix}  \\ \nonumber
     & \exp\left[\frac{j2\pi(\hat{r}^T_{rx,m}. \Bar{d}_{rx,u})}{\lambda_{o}}\right]
     \exp\left[\frac{j2\pi(\hat{r}^T_{tx,m}. \Bar{d}_{tx,s})}{\lambda_{o}}\right],\\ \nonumber
    %
\end{align}
where $\alpha_{m}$ is the received amplitude of the $m^{th}$ MP. The XPD ratios, $K_{m_{\theta,\phi}}$, $K_{m_{\phi,\theta}}$, $K_{m_{\phi,\phi}}$ are in linear scale and computed using NYUSIM \cite{nyusim}. Furthermore, {$\phi_{m}^{\theta\theta}, \phi_{m}^{\theta\phi}, \phi_{m}^{\phi\theta}, \phi_{m}^{\phi\phi}$} are generated for each MP $m$ in the interval $(0,2\pi)$. A detailed description of the remaining parameters in \eqref{nyuChannelCoefficient} is presented in \cite{3GPPTR}.
\subsection{Computation of PSD}
The computation of the received PSD in NYUSIM is similar to \cite{zugno2020implementation} and is given in \eqref{PSD},
\vspace{-0.1in}
\begin{align}
    \label{PSD}
    S_{rx}(t,f) = S_{tx}(t,f)w_{rx}^TH(t,f)w_{tx},
\end{align}
where $S_{tx}(t,f)$ denotes the PSD of the transmitted signal, $w_{rx}$ and $w_{tx}$ denote the Rx and Tx beamforming vectors, respectively, and $H(t,f)$ represents the channel matrix in the frequency domain. Taking the Fourier transform of the channel matrix \eqref{nyuChannelMatrix} and the channel coefficients in \eqref{nyuChannelCoefficient}, \eqref{PSD} can be mathematically expressed as 
\begin{align}
    \label{PSDFT}
    S_{rx}(t,f) &= S_{tx}(t,f)\sum_{n=1}^{N}\sum_{s=1}^{S}\sum_{u=1}^{U}w_{rx,u}H_{u,s,n}w_{tx,s}e^{j2\pi v_{n}t}e^{j2\pi \tau_{n}f} \nonumber \\
    &= S_{tx}(t,f)\sum_{n=1}^{N}L_{n}e^{j2\pi v_{n}t}e^{j2\pi \tau_{n}f},
\end{align}
where $N$ denotes the total number of MPCs, $S$ and $U$ represent the number of Tx and Rx antennas for MIMO, $v_{n}$ and $\tau_{n}$  denote the Doppler spread and delay of the $n^{th}$ MP. $L_{n}$ is called the long-term component of MP $n$ and is defined as 
\begin{align}
    \label{LongTerm}
    L_{n} = \sum_{s=1}^{S}\sum_{u=1}^{U}w_{rx,u}H_{u,s,n}w_{tx,s}. 
\end{align}
\texttt{NYU\-Spectrum\-Propagation\-Loss\-Model} extends the \texttt{Spectrum\-Prop\-agation\-Loss\-Model}. The channel coefficients are obtained from \texttt{NYU\-Channel\-Model} by the class \texttt{NYU\-Spectrum\-Propagation\-Loss\-M\-odel}. The functionality of the method \texttt{Do\-Calc\-Rx\-Power\-Spectral\-Density} is to compute \eqref{PSDFT}. The implementation and working of all the methods, namely  \texttt{Do\-Calc\-Rx\-Power\-Spectral\-Density}, \texttt{Get\-Long\-Term} and \texttt{Calc\-Beam\-forming\-Gain} are similar to the implementation of 3GPP SCM, details of which can be found in Section 3.5 of \cite{zugno2020implementation}.

\renewcommand{\arraystretch}{1.25}
\begin{table*}[]
  \centering
  \caption{\label{Distributions} Distributions Used in NYUSIM for the Frequency Range of 0.5-150 GHz in 3GPP-listed Scenarios}
  \resizebox{1.7\columnwidth}{!}{%
    \begin{tabular}{|N|M|c|c|c|c|c|}
    \hline
    \multicolumn{1}{|c|}{\textbf{Step}} & \multicolumn{1}{M|}{\textbf{Channel Parameters}} & \multicolumn{1}{c|}{\textbf{UMi}} & \multicolumn{1}{c|}{\textbf{UMa}} & \multicolumn{1}{c|}{\textbf{RMa}} & \multicolumn{1}{c|}{\textbf{InH}} & \textbf{InF} \\
    \hline
    \multicolumn{1}{|N|}{\multirow{3}[2]{*}{\textbf{1}}} & \multicolumn{1}{c|}{\multirow{3}[2]{3cm}{\centering Number of time clusters ($N$) \newline{}}} & \multicolumn{3}{c|}{\multirow{2}[1]{*}{$N\sim$DU(1,$N_c$)}} & \multicolumn{2}{c|}{\multirow{3}[2]{*}{$N \sim$ Poisson$(\lambda_{c}) + 1$}} \\
    \multicolumn{1}{|c|}{} &       & \multicolumn{3}{c|}{} & \multicolumn{2}{c|}{} \\
    \multicolumn{1}{|c|}{} &       & \multicolumn{3}{c|}{*DU: Discrete Uniform} & \multicolumn{2}{c|}{} \\
    \hline
    \multirow{4}[4]{*}{\textbf{2}} & \multirow{4}[4]{3cm}{\centering Number of cluster subpaths ($M_n$) } & \multicolumn{2}{l|}{$<$ 100 GHz} & \multirow{4}[4]{*}{$M_n$ $\sim$ DU(1, $M_s$)} & \multicolumn{2}{c|}{\multirow{2}[2]{*}{$M_n \sim (1-\beta)\delta$ ($M_n$) + DE($\mu_s$)}} \\
          &       & \multicolumn{2}{c|}{$M_n$ $\sim$ DU(1, $M_s$)} &       & \multicolumn{2}{c|}{} \\
\cline{3-4}          &       & \multicolumn{2}{l|}{$\geq$ 100 GHz} &       & \multicolumn{2}{c|}{\multirow{2}[2]{*}{*DE: Discrete Exponential}} \\
          &       & \multicolumn{2}{c|}{$Mn$ $\sim$ Exp($\mu_s$)} &       & \multicolumn{2}{c|}{} \\
    \hline
    \multirow{3}[4]{*}{\textbf{3}} & \multirow{3}[4]{3cm}{\centering Cluster excess delay $\tau_n$ (ns)} & \multicolumn{5}{c|}{$\tau_n = \tau_{n-1}+\rho_{M_{n-1},n-1}+\Delta\tau_n+\textup{MTI},$} \\
\cline{3-7}          &       & \multicolumn{3}{c|}{$\Delta\tau_n\sim \textup{Exp}(\mu_\tau)$} & $\Delta\tau_n\sim \textup{Exp}(\mu_\tau)\; \mathrm{or}\; \textup{Logn}(\mu_\tau,\sigma_\tau)$  & $\Delta\tau_n\sim \textup{Gamma}(\alpha_\tau,\beta_\tau)$ \\
          &       & \multicolumn{3}{c|}{MTI = 25 ns} & MTI = 6 ns & MTI = 8 ns \\
    \hline
    \multirow{5}[3]{*}{4} & \multirow{5}[3]{3cm}{\centering Intra-cluster subpath excess delay $\rho_{m,n}$ (ns)} & \multicolumn{3}{l|}{$<$ 100 GHz} & \multirow{5}[3]{*}{$\rho_{m,n} \sim \textup{Exp}(\mu_\rho)$} & \multirow{5}[3]{*}{$\rho_{m,n} \sim \textup{Gamma}(\alpha_\rho,\beta_\rho)$} \\
          &       & \multicolumn{3}{c|}{$\rho_{m,n}=\left[\frac{1}{B_{bb}}\times (m-1)\right]^{1+X_n}$,} &       &  \\
          &       & \multicolumn{3}{c|}{$m$ = 1,2,..$M_n$ $n$ = 1,2,..,$N$} &       &  \\
\cline{3-5}          &       & \multicolumn{3}{l|}{$\geq$ 100 GHz} &       &  \\
          &       & \multicolumn{3}{c|}{$\rho_{m,n} \sim \textup{Exp}(\mu_\rho)$} &       &  \\
    \hline
    \multirow{3}[2]{*}{\textbf{5}} & \multirow{3}[2]{3cm}{\centering Cluster power $P_n$ (mW)} & \multicolumn{5}{c|}{$P'_n = \Bar{P_0} e^{-\frac{\tau_n}{\Gamma}}10^{\frac{Z_n}{10}}$} \\
          &       & \multicolumn{5}{c|}{$P_n = \frac{P'_n}{\sum_{k=1}^{N}P'_k}\times P_r \textup{[mW]}$} \\
          &       & \multicolumn{5}{c|}{$Z_n \sim \mathcal{N}(0,\sigma_Z),\;n=1,2,...,N $} \\
    \hline
    \multirow{3}[2]{*}{\textbf{6}} & \multirow{3}[2]{3cm}{\centering Subpath power $\Pi_{m,n}$ (mW)} & \multicolumn{5}{c|}{$\Pi'_{m,n} = \bar{\Pi}_0 e^{-\frac{\rho_{m,n}}{\gamma}}10^{\frac{U_{m,n}}{10}}$} \\
          &       & \multicolumn{5}{c|}{$\Pi_{m,n} = \frac{\Pi'_{m,n}}{\sum_{k=1}^{M_n}\Pi'_k}\times P_n \textup{[mW]}$} \\
          &       & \multicolumn{5}{c|}{$U_{m,n} \sim \mathcal{N}(0,\sigma_U),\; m=1,2,...,M_n$} \\
    \hline
    \textbf{7}     & \centering Subpath phase $\varphi$ (rad) & \multicolumn{5}{c|}{Uniform(0,$2\pi$)} \\
    \hline
    \multirow{2}[2]{*}{\textbf{8}} & \multirow{2}[2]{3cm}{\centering Number of spatial lobes ($L_{AOD},L_{AOA}$)} & \multicolumn{3}{c|}{\multirow{2}[2]{*}{$L \sim \textup{Poisson}(\lambda_l) +1$}} & \multicolumn{1}{c|}{\multirow{2}[2]{*}{$L \sim \textup{DU}(1,\lambda_l)$}} & \multirow{2}[2]{*}{$L \sim \textup{Poisson}(\lambda_l) +1$} \\
          &       & \multicolumn{3}{c|}{} & \multicolumn{1}{c|}{} &  \\
    \hline
    \multirow{2}[2]{*}{\textbf{9}} & \multirow{2}[2]{3cm}{\centering Mean lobe azimuth angle $\phi_{i}(^o)$} & \multicolumn{5}{c|}{$\phi_i\sim \textup{Uniform}(\phi_{\textup{min}},\phi_{\textup{max}})$} \\
          &       & \multicolumn{5}{c|}{$\phi_{\textup{min}} = \frac{360(i-1)}{L}, \phi_{\textup{max}}=\frac{360i}{L},\;i=1,2,...,L$} \\
    \hline
    \textbf{10} & \centering Mean lobe elevation angle $\theta_i(^o)$ & \multicolumn{5}{c|}{$\theta_i \sim \textup{Gaussian}(\mu_l,\sigma_l)$} \\
    \hline
    \multirow{4}[1]{*}{\textbf{11}} & \multirow{4}[1]{3cm}{\centering Subpath angular offset from ($\Delta\phi,\Delta\theta$) from mean lobe angle $\phi_i,\theta_i$} & \multicolumn{5}{c|}{($\Delta\phi_i)_{m,n,AOD} \sim \mathcal{N}(0,\sigma_{\phi,AOD})$} \\
          &       & \multicolumn{5}{c|}{($\Delta\theta_i)_{m,n,ZOD} \sim \mathcal{N}(0,\sigma_{\theta,ZOD})$} \\
          &       & \multicolumn{5}{c|}{($\Delta\phi_j)_{m,n,AOA} \sim \mathcal{N}(0,\sigma_{\phi,AOA})$ } \\
          &       & \multicolumn{5}{c|}{$(\Delta\theta_j)_{m,n,ZOA} \sim \mathcal{N}(0,\sigma_{\theta,ZOA})$} \\
          \hline
    \end{tabular}%
    }
  \label{tab:distributions}%
\end{table*}%

\section{Example of using NYUSIM in ns-3}
The example file \texttt{nyu\--cha\-nnel\--exa\-mple.cc} included in the \texttt{Spe\-ctr\-um} module shows a sample usage of the NYUSIM wireless channel in ns-3 and it can be used with mmWave~\cite{mezzavilla2018end} in ns-3 to perform end-to-end network performance for cross-layer design. However, for illustration purposes, in this example, we only compute the mean path loss value. The script \texttt{nyu\--cha\-nnel\--exa\-mple.cc} is run once for 10 seconds for each scenario and channel condition, assuming a base station height of 35 m. The average mean path loss value is computed over 10 seconds. Figure \ref{fig:PathLoss} shows the propagation loss vs. distance using the script \texttt{nyu\--channel-\-example.cc} for all scenarios in LOS and NLOS channel conditions at 142 GHz. 
\\The procedure used to create and configure the channel model classes, assuming the UMa scenario is selected, is:
\begin{enumerate}
    \vspace{-0.08in}
    \item A \texttt{NYUUmaChannelConditionModel} instance is created in the first step.
    \item Create a \texttt{NYUUmaPropagationLossModel} instance, set the channel condition model using the \texttt{``Channel\-Condition\-Model''} attribute, and configure the carrier frequency using the \texttt{``Frequency''} attribute. 
    \item The UMa scenario, carrier frequency, bandwidth, and channel condition model are set using the attributes \texttt{``Scenario''}, \texttt{``Frequency''}, \texttt{``RfBandwidth''} and \texttt{``Channel\-Condition\-Mo\-del''} when creating an instance of the \texttt{NYU\-Spectrum\-Prop\-agation\-Loss\-Model}; 
    \item Create a \texttt{Three\-Gpp\-Antenna\-Array\-Model} instance for each device, then call the method \texttt{NYU\-Spectrum\-Propagation\-Loss\-Model::\-AddDevice} to inform the \texttt{NYU\-Spectrum\-Propagat\-ion\-Loss\-Model} class of the device-antenna associations. 
\end{enumerate}
In addition, a detailed comparison of full-stack end-to-end network simulations at 28 GHz using 3GPP SCM and NYUSIM for a single user scenario is presented in \cite{hp2023icc}. The other use cases for NYUSIM are similar to those stated for 3GPP SCM in \cite{zugno2020implementation}.

\section{Conclusions and Future Work}
\vspace{-0.1in}
In this article, we presented our implementation of the drop-based NYUSIM in ns-3 to run Monte Carlo simulation \cite{tranter2004principles} to evaluate the performance of modems and end-to-end wireless systems across all layers of the protocol stack \cite{hp2023icc}. This work will enable the research community to investigate the impact of wireless channels on system/network performance in mmWave and sub-THz frequencies using real-world measurement-based channel models for 3GPP-listed scenarios. In Section 1, we underscored the limitation of the 3GPP SCM and the importance of using accurate measurement-based channel models to design future wireless systems. Section 2 described the implementation details of NYUSIM in ns-3. Finally, in Section 3, we listed the examples and use cases for NYUSIM. In the future, we plan to implement NYUSIM-based blockage models for all 3GPP SCM listed scenarios and create LOS probability models for InH, RMa, and InF scenarios. In addition, we will also add spatial consistency-based models for NYUSIM in ns-3 and extend NYUSIM models above 150 GHz for 3GPP-listed scenarios.

\begin{acks}
We would like to thank Tom Henderson for his support. This work is supported by the NYU
WIRELESS industrial affiliates program, and the commissioned research (No.04201) from the National Institute of Information and Communications Technology (NICT), Japan.
\end{acks}
\balance

\end{document}